\documentclass[12pt,a4paper]{article}
\usepackage{graphicx}
\usepackage[T1]{fontenc}
\usepackage[utf8]{inputenc}
\usepackage{textcomp}
\usepackage[sc,osf]{mathpazo}
\usepackage{a4wide}  
\usepackage{latexsym,amsthm,amsfonts,amsmath,mathrsfs,amssymb}
\usepackage{dsfont}
\usepackage{accents}
\usepackage[nosort]{cite}
\usepackage{booktabs} 
\usepackage[unicode,implicit]{hyperref}
\hypersetup{%
  pdftitle    = {On scalar charges and black-hole thermodynamics}
  pdfkeywords = {Black holes, thermodynamics, Noether charge, Wald entropy,
    Komar integral, Smarr formula, first law, scalar charges, conservation},
  pdfauthor   = {Romina Ballesteros, Carmen G\'omez-Fayr\'en, Tom\'as Ort\'{\i}n and Matteo Zatti},
  plainpages  = true,
  colorlinks  = true,
  citecolor   = blue,
  urlcolor    = red,
  linkcolor   = black
}
\newcommand{\hepth}[1]{{\tt
\href{http://www.arXiv.org/abs/hep-th/#1}{hep-th/#1}}}
\newcommand{\grqc}[1]{{\tt
\href{http://www.arXiv.org/abs/gr-qc/#1}{gr-qc/#1}}}

\newcommand{\arxiv}[1]{{\tt arXiv:\href{http://www.arXiv.org/abs/#1}{#1}}}
\newcommand{\doi}[1]{{\tt DOI:\href{http://dx.doi.org/#1}{#1}}}

\allowdisplaybreaks

\makeatletter
\@addtoreset{equation}{section}
\makeatother

\begin{document}

\begin{flushright}
\small
IFT-UAM/CSIC-23-018\\
\texttt{arXiv:2302.11630  [hep-th]}\\
February 23\textsuperscript{rd}, 2023\\
\normalsize
\end{flushright}

\vspace{1cm}

\begin{center}

  {\Large {\bf On scalar charges and black hole thermodynamics}}

\vspace{1.5cm}

\renewcommand{\thefootnote}{\alph{footnote}}

{\sl\large Romina Ballesteros,$^{1,2,}$}\footnote{Email: {\tt romina.ballesteros[at]estudiante.uam.es}}
{\sl\large Carmen G\'omez-Fayr\'en,$^{1,}$}\footnote{Email: {\tt carmen.gomez-fayren[at]estudiante.uam.es}}
{\sl\large Tom\'{a}s Ort\'{\i}n$^{1,}$}\footnote{Email: {\tt tomas.ortin[at]csic.es}}\\
{\sl\large and Matteo Zatti$^{1,}$}\footnote{Email: {\tt matteo.zatti[at]estudiante.uam.es}}

\setcounter{footnote}{0}
\renewcommand{\thefootnote}{\arabic{footnote}}

\vspace{1cm}

$^{1}${\it Instituto de F\'{\i}sica Te\'orica UAM/CSIC\\
C/ Nicol\'as Cabrera, 13--15,  C.U.~Cantoblanco, E-28049 Madrid, Spain}

\vspace{.5cm}

$^{2}${\it Pontificia Universidad Cat\'olica de Valpara\'{\i}so,
  Instituto de F\'{\i}sica, Av. Brasil 2950, Valpara\'{\i}so, Chile}

\setcounter{footnote}{0}
\renewcommand{\thefootnote}{\arabic{footnote}}
\vspace{1cm}

\vspace{1cm}


{\bf Abstract}
\end{center}
\begin{quotation}
  {\small We revisit the first law of black hole thermodynamics in
    4-dimensional theories containing scalar and Abelian vector fields coupled
    to gravity using Wald's formalism and a new definition of scalar charge as
    an integral over a 2-surface which satisfies a Gauss law in the background
    of stationary black-hole spacetimes. We focus on ungauged
    supergravity-inspired theories with symmetric sigma models whose
    symmetries generate electric-magnetic dualities leaving invariant their
    equations of motion. Our manifestly duality-invariant form of the first
    law is compatible with the one obtained by of Gibbons, Kallosh and Kol. We
    also obtain the general expression for the scalar charges of a stationary
    black hole in terms of the other physical parameters of the solution and
    the position of the horizon, generalizing the expression obtained by
    Pacilio for dilaton black holes. }
\end{quotation}

\newpage
\pagestyle{plain}



\section*{Introduction}

It is widely believed that one of the defining characteristics of classical
black holes is that they have no ``hair''. The concept of black-hole hair is a
very broad one but, for the stationary black holes we will be concerned with
in this paper it can be defined as any parameter that enters the metric and
which cannot be eliminated through a coordinate transformation which is not a
function of the charges of the theory which are conserved by virtue of a local
symmetry (mass, angular momenta, electric charges) or a topological property
(magnetic charges) or the asymptotic values of the scalars (\textit{moduli}).

Scalar charges, typically defined through the asymptotic behavior at spatial
infinity of the scalars in the black-hole spacetime, are not protected by any
conservation law. In ungauged theories the only local symmetries scalar fields
transform under are diffeomorphisms but the conserved charges associated to
them are the gravitational ones: mass and linear and angular momenta. Scalar
fields only transform under global symmetries of the action or of the
equations of motion only to which we will refer to as dualities. However, the
charges associated to those symmetries in stationary black-hole spacetimes
vanish identically. They seem to have nothing to do with the
conventionally-defined black-hole scalar charges. Gauging the global
symmetries does not help because the gauge symmetry would be associated to
some 1-form gauge fields and the conserved charges would have the
interpretation of electric and magnetic charges. 

Therefore, according to our definition of hair, scalar charges are understood
as hair and, according to the \textit{no-hair conjecture}, no black-hole
solutions with regular horizons (henceforth to be referred to as ``regular
black holes'') carrying scalar charges should be expected. Any scalar charges
possessed by gravitationally collapsing matter should be radiated away in the
black-hole formation. However, there are many regular black hole solutions
carrying non-vanishing scalar charges such as dilaton black holes and their
generalizations.\footnote{For a review with many references, see
  Ref.~\cite{Ortin:2015hya}.}

The solution to this apparent counterexample of the no-hair conjecture lies
in the distinction between primary and secondary hair \cite{Coleman:1991ku}:
in all the regular black-hole solutions with non-vanishing scalar charges,
those charges are not independent parameters but very specific functions of
the independent conserved charges which are allowed by the no-hair conjecture
and they are (by definition) secondary hair. In the solutions in which the
scalar charges are truly independent parameters, such as the
Janis-Newman-Winicour solution \cite{Janis:1968zz} or the Agnese-La Camera
solutions \cite{Agnese:1994zx} and their generalizations \cite{Ortin:2015hya},
there are no regular horizons but naked singularities unless the scalar charge
takes the value of the specific function of the conserved charges we mentioned
above (simply zero in the JNW solution). This kind of scalar hair is, by
definition, primary hair and it is the one which would actually be forbidden
by the conjecture.

The scalar charges which are allowed by the no-hair conjecture remain,
nevertheless, quite mysterious: What are the values of the scalar charges
allowed in a given theory?  Why are those values allowed and no others? And,
even more basic: Is there a coordinate-independent definition of scalar
charge?

This mystery only deepened when Gibbons, Kallosh and Kol (GKK) showed in
Ref.~\cite{Gibbons:1996af} (see also Ref.~\cite{Astefanesei:2018vga}) that the
allowed scalar charges occur in the first law of black hole mechanics
\cite{Bardeen:1973gs} as thermodynamical potentials conjugate to the
variations of the moduli. While it is not clear which kind of physical process
may result in a change of the moduli,\footnote{The same could be said about
  magnetic charges.} it is a fact that varying the black-hole entropy formulae
of known solutions with respect to the moduli one finds the scalar charges as
coefficients of those variations.

Wald's formalism \cite{Lee:1990nz,Wald:1993nt,Iyer:1994ys} opened a new venue
for the study of black-hole thermodynamics that can be used to explore the
role of scalar charges into it. The main observation, realized in the context
of purely gravitational (matter-free) theories invariant under diffeomorphisms
is that the properties of the Noether $(d-2)$-form charge associated to the
invariance under diffeomorphisms (\textit{Noether-Wald charge}) can be used to
prove the first law of black-hole thermodynamics.

In theories with matter, this law includes work terms proportional to the
variations of conserved charges and the GKK scalar term proportional to the
variations of the moduli.  In the last few years we have extended the
formalism to handle theories in which there are matter fields with gauge
symmetries coupled to gravity showing how the electric work terms appear
\cite{Elgood:2020svt,Elgood:2020mdx,Elgood:2020nls},\footnote{A slightly
  different approach to the one taken in those papers, which is the one used
  here as well, is the point of view of ``invariance up to gauge
  transformations'', taken in
  Refs.~\cite{Sudarsky:1992ty,Sudarsky:1993kh,Barnich:2007bf,McCormick:2013nkb,Hajian:2013lna,Hajian:2015xlp}.}
showing how extended black-hole thermodynamics arises in this context
\cite{Ortin:2021ade,Meessen:2022hcg}, how to include magnetic charges in the
first law \cite{Ortin:2022uxa} and how to construct Komar integrals from which
Smarr formulae can be derived \cite{Ortin:2021ade,Mitsios:2021zrn}. In all
those cases each new work term in the first laws is associated to a gauge
symmetry or an equivalent topological property. Since, as we have seen, scalar
charges are not associated to neither, it is unclear how the GKK work term can
be recovered in Wald's formalism.\footnote{In extended thermodynamics there
  are work terms associated to the variation of dimensionful constants which,
  apparently, unrelated to gauge symmetries.\footnote{This general fact was
    conjectured in Ref.~\cite{Hajian:2021hje}. The Regge slope parameter
    $\alpha'$ that controls higher order corrections in the string effective
    action has dimensions of length squared and the occurrence of term
    proportional to its variations in the first law of $\alpha'$-corrected
    black holes has been demonstrated in Ref.~\cite{Cano:2022tmn}.} However,
  those constants can be dualized into $(d-1)$-form potentials with a gauge
  freedom (for the cosmological constant, see
  Refs.~\cite{Freund:1980xh,Aurilia:1980xj}) and this description leads to the
  work terms
  \cite{Chernyavsky:2017xwm,Hajian:2016iyp,Hajian:2021hje,Ortin:2021ade,Meessen:2022hcg}.}
The absence of a good coordinate-independent definition for the scalar charge
complicates this problem.

In this paper we are going to show how this problem can be solved taking into
account hitherto ignored contributions to the integrals at spatial infinity and
using a definition of scalar charge as the integral of a $(d-2)$-form which is
manifestly coordinate and gauge independent and which satisfies a Gauss law in
stationary black-hole spacetimes. This definition relies in the existence of
conserved charges associated to global symmetries and in the existence of a
timelike Killing vector whose Killing horizon coincides with the black-hole's
event horizon and whose action leaves invariant all the physical
fields. Therefore, there is a scalar charge associated to each global
symmetry, and, therefore, the number of charges may or may not coincide with
the number of scalar fields.

In this paper we have studied 4-dimensional theories\footnote{The extension to
  higher dimensions and higher-rank forms is straightforward using the results
  of Ref.~\cite{Bandos:2016smv} for the Noether-Gaillard-Zumino currents.}
whose scalar kinetic terms are described by symmetric sigma models in which
the scalar fields map spacetime into a target space which is a symmetric
Riemannian homogeneous space G/H. These kinetic terms are very common in
supergravity theories. Furthermore, our theories include Abelian 1-forms and
we are going to assume that the couplings of the scalars to those 1-forms are
such that the equations of motion, enhanced with the Bianchi identities
satisfied by the 2-form field strengths are invariant under the duality group
G.\footnote{The general form of the theories that we consider is identically to
  that of the theories considered by GKK in Ref.~\cite{Gibbons:1996af} but, in
  our approach it is crucial to know the global symmetries of the theory.}
Again, this is a fairly common situation in supergravity and include simple
theories such as the Einstein-Maxwell-Dilaton ones. In these theories we can
associate a conserved scalar charge to each of the generators of G, even if
some of the transformations (the electric-magnetic duality rotations in
particular) do not leave the action invariant. As a result, according to our
definition, there are always more scalar charges than scalars. Nevertheless,
we are going to show that the conventional scalar charges can be recovered as
combinations of the ones we have defined and we are going to check these
relations in particular black-hole solutions.

In this framework we are going to proof the first law of black-hole
thermodynamics recovering the GKK results and, as a byproduct, we are going to
find a general expression for the scalar charges in terms of the conserved
charges and the position of the horizon, thus answering one of the
long-standing questions posed above.\footnote{After completion of this work,
  we found that a similar definition of scalar charge and similar result had
  been found in Ref.~\cite{Pacilio:2018gom} in the context of the dilaton
  black holes of the EMD theories.} Observe that, since our
definition of scalar charge satisfies a Gauss Law, the value obtained for
those charges is the same whether we calculate the integrals over the horizon
or at infinity.

This paper is organized as follows: in Section~\ref{sec-theory} we review the
kind of theories that we are considering, their duality symmetries, the
Gaillard-Zumino theorem \cite{Gaillard:1981rj} and the construction of the
\textit{Noether-Gaillard-Zumino} (NGZ) currents which will be used in
Section~\ref{sec-scalarcharge} to define the scalar charges. In
Section~\ref{sec-firstlaw} we derive the first law recovering the GKK results
and the general expression of scalar charges in terms of conserved charges and
the position of the horizon. In Sections~\ref{sec-staticdilatonBHs}
and~\ref{sec-staticBHs} we test our results on dilaton and axion-dilaton black
holes respectively. Section~\ref{sec-discussion} contains a discussion of our
results.

\section{The theory}
\label{sec-theory}

In this section we are going to review the theories we are going to consider
and their duality symmetries. Most of this material can be found elsewhere,
but here we adapt it to our needs and conventions.

Throughout this paper we are going to consider 4-dimensional ungauged
supergravity-inspired theories containing $n_{S}$ scalar fields $\phi^{x}$
that parametrize a symmetric coset space $G/H$ and $n_{V}$ 1-form fields
$A^{\Lambda}=A^{\Lambda}{}_{\mu}dx^{\mu}$ with 2-form field strengths

\begin{equation}
F^{\Lambda} = dA^{\Lambda}\,,
\end{equation}

\noindent
coupled to gravity which we will describe through the Vierbein
$e^{a}=e^{a}{}_{\mu}dx^{\mu}$. Up to two derivatives, they can be described by
the generic action

\begin{equation}
\label{eq:action}
\begin{aligned}
  S
  & =
\frac{1}{16\pi G_{N}^{(4)}}\int 
  \left[ -\star (e^{a}\wedge e^{b}) \wedge R_{ab}
    +\tfrac{1}{2}g_{xy}d\phi^{x}\wedge \star d\phi^{y}
    -\tfrac{1}{2}I_{\Lambda\Sigma}F^{\Lambda}\wedge \star F^{\Sigma}
    -\tfrac{1}{2}R_{\Lambda\Sigma}F^{\Lambda}\wedge F^{\Sigma}
\right]
\\
& \\
& \equiv
\int \mathbf{L}\,,
\end{aligned}
\end{equation}

\noindent
where the kinetic matrix $I=\left(I_{\Lambda\Sigma}\right)$ is
negative-definite and we are going to assume that the positive-definite
$\sigma$-model metric $g_{xy}(\phi)$ is invariant under the action of $G$ (the
duality group) which also leaves invariant the set of all equations of motion
plus the Bianchi identities of the theory. This assumption will be translated
into conditions for the scalar-dependent matrices
$I=\left(I_{\Lambda\Sigma}\right)$ and $R= \left(R_{\Lambda\Sigma}\right)$
shortly.

We will set $G_{N}^{(4)}=1$ and we will ignore the normalization factor
$(16\pi)^{-1}$ for the time being.

The equations of motion are defined by (here $\varphi$ stands for all the
fields of the theory)

\begin{equation}
  \delta S
  =
  \int\left\{
    \mathbf{E}_{a}\wedge \delta e^{a} + \mathbf{E}_{x}\delta\phi^{x}
    +\mathbf{E}_{\Lambda}\delta A^{\Lambda}  
    +d\mathbf{\Theta}(\varphi,\delta\varphi)
    \right\}\,,
\end{equation}

\noindent
and given by

\begin{subequations}
  \begin{align}
    \label{eq:Ea}
    \mathbf{E}_{a}
    & =
      \imath_{a}\star(e^{b}\wedge e^{c})\wedge R_{bc}
      +\tfrac{1}{2}g_{xy}\left(\imath_{a}d\phi^{x} \star d\phi^{y}
      +d\phi^{x}\wedge \imath_{a}\star d\phi^{y}\right)
      \nonumber \\
    & \nonumber \\
    & \hspace{.5cm}
      -\tfrac{1}{2}I_{\Lambda\Sigma}\left(\imath_{a}F^{\Lambda}\wedge\star
      F^{\Sigma} -F^{\Lambda}\wedge \imath_{a}\star F^{\Sigma}\right)\,,
    \\
    & \nonumber \\
    \mathbf{E}_{x}
    & =
      -g_{xy}\left\{d\star d\phi^{y}
      +\Gamma_{zw}{}^{y}d\phi^{z}\wedge\star d\phi^{w} \right\}
      -\tfrac{1}{2}\partial_{x}I_{\Lambda\Sigma} F^{\Lambda}\wedge\star
      F^{\Sigma}
      -\tfrac{1}{2}\partial_{x}R_{\Lambda\Sigma} F^{\Lambda}\wedge F^{\Sigma}\,,
    \\
    & \nonumber \\
    \mathbf{E}_{\Lambda}
    & =
      d F_{\Lambda}\,,
  \end{align}
\end{subequations}

\noindent
where we have defined the dual 2-form field strength

\begin{equation}
  \label{eq:dualfieldstrengthsdef}
  F_{\Lambda}
  \equiv
  I_{\Lambda\Sigma}\star F^{\Sigma}+R_{\Lambda\Sigma}F^{\Sigma}\,.
\end{equation}

Furthermore,

\begin{equation}
  \label{eq:Theta}
  \mathbf{\Theta}(\varphi,\delta\varphi)
  =
  -\star (e^{a}\wedge e^{b})\wedge \delta \omega_{ab}
  +g_{xy}\star d\phi^{x}\delta\phi^{y}
  -F_{\Lambda}\wedge \delta A^{\Lambda}\,.
\end{equation}

The original and dual 2-forms can be combined into a symplectic vector of
2-forms\footnote{The symplectic nature of this vector will be proven shortly.}

\begin{equation}
  \left(F^{M}\right)
  \equiv
  \left(
    \begin{array}{c}
      F^{\Lambda} \\ F_{\Lambda} \\
    \end{array}
    \right)\,,
\end{equation}

\noindent
and the Bianchi identities of the original 2-form field strength $F^{\Lambda}$

\begin{equation}
dF^{\Lambda}=0\,,  
\end{equation}

\noindent
and  the Maxwell equations $\mathbf{E}_{\Lambda}=0$ can be written as

\begin{equation}
  \label{eq:equations}
dF^{M}=0\,.  
\end{equation}

These equations can be interpreted as Bianchi identities implying the local
existence of 1-form potentials 

\begin{equation}
  \left(A^{M}\right)
  \equiv
  \left(
    \begin{array}{c}
      A^{\Lambda} \\ A_{\Lambda} \\
    \end{array}
    \right)\,,
\end{equation}

\noindent
such that

\begin{equation}
F^{M}=dA^{M}\,.  
\end{equation}

The set of equations (\ref{eq:equations}) is invariant under arbitrary
GL$(2n_{V},\mathbb{R})$ transformations

\begin{equation}
F^{M\,\prime}  =S^{M}{}_{N}F^{N}\,,
\end{equation}

\noindent
but we have to take into account the rest of the equations and an important
constraint: the components of $F^{M}$ are not independent and, therefore,
$F^{M}$ satisfies the following \textit{twisted self-duality constraint}

\begin{equation}
  \label{eq:twisted}
\star F^{M}= -\Omega^{MN}\mathcal{M}_{NP}F^{P}\,,  
\end{equation}

\noindent
where $\mathcal{M}_{MN}$ is the $2n_{V}\times 2n_{V}$ symmetric symplectic
matrix


\begin{equation}
  \begin{aligned}
    \mathcal{M}
    & =
    \left(\mathcal{M}_{MN}\right)
    =
    \left(
      \begin{array}{lr}
        I +RI^{-1}R
        &
          -RI^{-1}
        \\
        & \\
        -I^{-1}R
        &
          I^{-1}
        \\
      \end{array}
    \right)\,,
    \\
    & \\
    \mathcal{M}^{-1}
    & =
    \left(\mathcal{M}^{MN}\right)
    =
    \Omega^{-1\, T}\mathcal{M}\Omega
    =
    \left(
      \begin{array}{lr}
        I^{-1}
        &
          I^{-1}R
        \\
        & \\
        RI^{-1}
        &
          I +RI^{-1}R
        \\
      \end{array}
    \right)\,,
  \end{aligned}
\end{equation}

\noindent
and

\begin{equation}
  \Omega
  =
  \left(\Omega_{MN}\right)
  = 
  \left(
    \begin{array}{lr}
      0 & \mathbb{1}_{n_{V}\times n_{V}}   \\
      & \\
-\mathbb{1}_{n_{V}\times n_{V}}      & 0 \\
    \end{array}
  \right)\,,
  \hspace{1cm}
  \Omega^{-1}
  = \left(\Omega^{PN}\right)\,.
\end{equation}

As a consequence, the set of Maxwell equations and Bianchi identities will
only be invariant under the subset of GL$(2n_{V},\mathbb{R})$ transformations
that preserve this constraint, which is possible provided that $\mathcal{M}$
transforms as

\begin{equation}
  \label{eq:Mtransform}
  \mathcal{M}'
  =
  \left(\Omega^{-1}S\Omega\right) \mathcal{M} S^{-1}\,,
  \hspace{1cm}
  S = \left(S^{M}{}_{N}\right)\,.
\end{equation}

It is convenient to analyze the invariance of the Einstein equations first.

Using the identity

\begin{equation}
\mathcal{M}_{MN}\imath_{a}F^{M}\wedge \star F^{N}  
=
I_{\Lambda\Sigma}\left(\imath_{a}F^{\Lambda}\wedge\star
      F^{\Sigma} -F^{\Lambda}\wedge \imath_{a}\star F^{\Sigma}\right)\,,
\end{equation}

\noindent
and the twisted self-duality constraint Eq.~(\ref{eq:twisted}), 
the energy-momentum tensor of the 1-forms can be written in the form

\begin{equation}
-\Omega_{MN}\imath_{a}\star F^{M}\wedge \star F^{N}\,,    
\end{equation}

\noindent
which is left invariant by the transformations that leave invariant $\Omega$

\begin{equation}
  \label{eq:symplecticdef}
S^{T}\Omega S =\Omega\,,  
\end{equation}

\noindent
that is, by transformations that belong to Sp$(2n_{V},\mathbb{R})$
\cite{Gaillard:1981rj}.  Defining the $n_{V}\times n_{V}$ blocks of the
symplectic matrix $S$

\begin{equation}
  S
  =
  \left(
    \begin{array}{cc}
     A & B \\ C & D\\
    \end{array}
  \right)\,,
\end{equation}

\noindent
the symplectic nature of $S$ implies the following conditions for them:

\begin{subequations}
  \begin{align}
    A^{T}C - C^{T}A
    & =
      0\,,
    \\
    & \nonumber \\
    B^{T}D - D^{T}B
    & =
      0\,,
    \\
    & \nonumber \\
    D^{T}A-B^{T}C
    & =
      \mathbb{1}_{n_{V}\times n_{V}}\,.
  \end{align}
\end{subequations}

It is not difficult to see that, if $S$ symplectic, so is $S^{T}$.  The
symplectic nature of $S^{T}$ implies

\begin{subequations}
  \begin{align}
    BA^{T} -AB^{T}
    & =
      0\,,
    \\
    & \nonumber \\
    DC^{T} - CD^{T}
    & =
      0\,,
    \\
    & \nonumber \\
    DA^{T}-CB^{T}
    & =
      \mathbb{1}_{n_{V}\times n_{V}}\,.
  \end{align}
\end{subequations}

On the other hand, Eq.~(\ref{eq:symplecticdef}) implies

\begin{equation}
\Omega^{-1}S\Omega  = S^{-1\, T}\,,
\end{equation}

\noindent
and, going back to Eq.~(\ref{eq:Mtransform}), we find that 
\begin{equation}
  \label{eq:Mtransform2}
  \mathcal{M}^{-1\, \prime}
  =
  S\mathcal{M}^{-1} S^{T}\,.
\end{equation}

Defining the $n_{V}\times n_{V}$, symmetric, \textit{period matrix}

\begin{equation}
\mathcal{N} = R+iI\,,  
\end{equation}

\noindent
it can be seen that the transformation of $\mathcal{M}$
Eq.~(\ref{eq:Mtransform2}) is equivalent to the following generalized
fractional-linear transformations of $\mathcal{N}$:

\begin{equation}
  \label{eq:Ntransform}
  \mathcal{N}'
  =
  \left(C+D\mathcal{N}\right) \left(A+B\mathcal{N}\right)^{-1}\,.
\end{equation}

It is clear that these transformations of the period matrix are associated to
transformations of the scalars which we are going to study in their
infinitesimal form. The transformations of the scalars that leave the
equations of motion invariant must necessarily be generated by the Killing
vectors of the $\sigma$-model metric $g_{xy}$, which we are going to denote by
$\{k_{A}{}^{x}(\phi)\}$.\footnote{These transformations leave exactly
  invariant the energy-momentum tensor of the scalars, which is the only piece
  of the Einstein equations that we had not studied and transform covariantly
  the first two terms of the scalar equations of motion:
  \begin{equation}
 \delta_{A} \left\{d\star d\phi^{x}
   +\Gamma_{yz}{}^{x}d\phi^{y}\wedge\star d\phi^{z} \right\}
 =
 \partial_{w}k_{A}^{x}\left\{d\star d\phi^{w}
      +\Gamma_{yz}{}^{w}d\phi^{y}\wedge\star d\phi^{z} \right\}\,.
  \end{equation}
} In some cases it is
convenient to include in this set some vectors which are identically zero so
that the index $A$ can be used to label also transformations of the 1-form
fields that do not involve the scalars, if necessary. Of course, additional
conditions involving the kinetic matrices (hence, the period matrix) need to
be satisfied.

The infinitesimal transformations of the 1-form fields are

\begin{equation}
  \begin{aligned}
    S
    & \sim
    \mathbb{1}_{2n_{V}\times 2n_{V}} +\alpha^{A}T_{A}\,,
    \\
    & \\
    T_{A} & =
    \left(T_{A}{}^{M}{}_{N}\right)
    = 
    \left(
      \begin{array}{cc}
        T_{A}{}^{\Lambda}{}_{\Sigma} &  T_{A}{}^{\Lambda\Sigma} \\
                                     & \\
        T_{A\, \Lambda\Sigma} &  T_{A\, \Lambda}{}^{\Sigma} \\        
      \end{array}
    \right)\,.
  \end{aligned}
\end{equation}

$S$ is symplectic if

\begin{equation}
  T_{A}{}^{T}\Omega +\Omega T_{A} = 0\,,
  \,\,\,\,\,
  \Rightarrow
  \left(\Omega T_{A}\right)^{T}
  =
  \Omega T_{A}\,,
\end{equation}

\noindent
(so $\Omega_{MP} T_{A}{}^{P}{}_{N}$ is symmetric in $MN$) which implies, for
the block matrices

\begin{equation}
  \begin{aligned}
    T_{A\, \Lambda\Sigma}
    & = 
    T_{A\, \Sigma\Lambda}\,,
    \\
    & \\
    T_{A}{}^{\Lambda}{}_{\Sigma}
    & =
    -T_{A}{}_{\Sigma}{}^{\Lambda}\,,
    \\
    & \\
    T_{A}{}^{\Lambda\Sigma}
    & = 
    T_{A}{}^{\Sigma\Lambda}\,.    
  \end{aligned}
\end{equation}

Then, the infinitesimal form of Eq.~(\ref{eq:Ntransform}) is

\begin{equation}
  \label{eq:infinitesimalfractionallinearN}
  \delta_{A}\mathcal{N}_{\Lambda\Sigma}
  =
  T_{A\, \Lambda\Sigma}
  +T_{A}{}_{\Lambda}{}^{\Omega}\mathcal{N}_{\Omega\Sigma}
  -\mathcal{N}_{\Lambda\Omega}T_{A}{}^{\Omega}{}_{\Sigma}
  -\mathcal{N}_{\Lambda\Gamma} T_{A}{}^{\Gamma\Omega}\mathcal{N}_{\Omega\Sigma}\,,
\end{equation}

\noindent
and, for the kinetic matrices,

\begin{subequations}
  \begin{align}
 \label{eq:equivarianceconditionR}
    \delta_{A}R_{\Lambda\Sigma}
    & =
  T_{A\, \Lambda\Sigma}
  +T_{A}{}_{\Lambda}{}^{\Omega}R_{\Omega\Sigma}
  -R_{\Lambda\Omega}T_{A}{}^{\Omega}{}_{\Sigma}
  -R_{\Lambda\Gamma} T_{A}{}^{\Gamma\Omega}R_{\Omega\Sigma}
  +I_{\Lambda\Gamma} T_{A}{}^{\Gamma\Omega}I_{\Omega\Sigma}\,,
    \\
    & \nonumber \\
 \label{eq:equivarianceconditionI}
    \delta_{A}I_{\Lambda\Sigma}
    & =
  T_{A}{}_{\Lambda}{}^{\Omega}I_{\Omega\Sigma}
  -I_{\Lambda\Omega}T_{A}{}^{\Omega}{}_{\Sigma}
  -2R_{(\Lambda|\Gamma} T_{A}{}^{\Gamma\Omega}I_{\Omega|\Sigma)}\,.
  \end{align}
\end{subequations}

Then, it can be easily seen that the whole scalar equations of motion
transform as

\begin{equation}
  \delta_{A}\mathbf{E}_{x}
  = -\partial_{x}k_{A}{}^{y}\mathbf{E}_{y}\,,
\end{equation}

\noindent
under the transformations

\begin{equation}
  \delta_{A}\phi^{x}= k_{A}{}^{x}\,,
  \hspace{1cm}
  \delta_{A}F^{M} = T_{A}{}^{M}{}_{N}F^{N}\,,
\end{equation}

\noindent
provided that

\begin{equation}
  \label{eq:equivariancecondition}
k_{A}{}^{x}\partial_{x}\mathcal{N} = \delta_{A}\mathcal{N}\,,  
\end{equation}

\noindent
where $\delta_{A}\mathcal{N}$ is the infinitesimal generalized
fractional-linear transformation in
Eq.~(\ref{eq:infinitesimalfractionallinearN}) (or equivalently, in
Eqs.~(\ref{eq:equivarianceconditionR}) and (\ref{eq:equivarianceconditionI})
for the kinetic matrices). This equivariance condition of the kinetic matrices
is the condition we announced when we defined the theory.

\section{A definition of scalar charge}
\label{sec-scalarcharge}

Not all the symmetries of the equations of motion that we have studied are
symmetries of the action: those generated by $T_{A}{}^{\Lambda\Sigma}$ do not
leave the action invariant. Those generated by $T_{A\, \Lambda\Sigma}$ leave
it invariant up to a total derivative. However, as shown in
Ref.~\cite{Gaillard:1981rj}, there is an on-shell conserved current for each
of them, the so-called \textit{Noether-Gaillard-Zumino (NGZ) current}. The
simplest way to construct them is by contracting the scalar equations of motion
with the Killing vectors that generate them. Using the Killing vector equation
and the equivariance conditions Eqs.~(\ref{eq:equivarianceconditionR}) and
(\ref{eq:equivarianceconditionI}) we get \cite{Bandos:2016smv}

\begin{equation}
  \label{eq:kAxEx}
  \begin{aligned}
    k_{A}{}^{x}\mathbf{E}_{x}
    & =
    -d \star \hat{k}_{A} -\tfrac{1}{2}\Omega_{MP}T_{A}{}^{P}{}_{N}F^{M}\wedge F^{N}
    \\
    & \\
    & =
    -d\left[ \star \hat{k}_{A}
      +\tfrac{1}{2}\Omega_{MP}T_{A}{}^{P}{}_{N}A^{M}\wedge F^{N}\right]
    +\tfrac{1}{2}\Omega_{MP}T_{A}{}^{P}{}_{N}A^{M}\wedge \mathbf{E}^{N}\,,
  \end{aligned}
\end{equation}

\noindent
where we have collected in a symplectic vector of 3-forms the Maxwell equations
and Bianchi identities:

\begin{equation}
  \left(\mathbf{E}^{M}\right)
\equiv
  \left(
    \begin{array}{c}
      \mathbf{E}^{\Lambda} \\ \mathbf{E}_{\Lambda} \\
    \end{array}
    \right)\,,
\end{equation}

\noindent
and where we have denoted by $\hat{k}_{A}$ the pullback of the 1-form dual to
the target space Killing vector $k_{A}$

\begin{equation}
  \hat{k}_{A}
  \equiv
  k_{A}{}^{x}g_{xy}d\phi^{y}\,.
\end{equation}

Therefore, we find that the NGZ currents

\begin{equation}
  \star j_{A}
  \equiv
  -\star \hat{k}_{A} -\tfrac{1}{2}\Omega_{MP}T_{A}{}^{P}{}_{N}A^{M}\wedge F^{N}\,,
\end{equation}

\noindent
are conserved on-shell

\begin{equation}
  d\star j_{A}
  =
  k_{A}{}^{x}\mathbf{E}_{x}
  -\tfrac{1}{2}\Omega_{MP}T_{A}{}^{P}{}_{N}A^{M}\wedge \mathbf{E}^{N}
  \doteq
  0\,.
\end{equation}
  
The conservation of these currents follows from a global symmetry and the
associated charges are expressed as integrals over spacelike hypersurfaces
(volumes)

\begin{equation}
q_{A} \sim \int_{\Sigma^{3}}\star j_{A}\,.  
\end{equation}

However, it is not difficult to see that in static black hole solutions with
non-trivial scalar fields $\phi^{x}$ whose charges $\Sigma^{x}$ are
conventionally defined\footnote{See, for instance,
  Ref.~\cite{Gibbons:1996af}.} through the asymptotic behavior of the field at
spatial infinity

\begin{equation}
  \label{eq:conventionaldefinition}
  \phi^{x} \stackrel{r\rightarrow \infty}{\sim}
  \phi^{x}_{\infty}+\frac{\Sigma^{x}}{r}\,,  
\end{equation}

\noindent
the NGZ charges not only do not reproduce the charges $\Sigma^{x}$ (or
combinations of them) but  vanish identically.

In stationary black hole spacetimes, though, there is another definition of
scalar charge that satisfies a Gauss law. Let us assume that all the fields
are invariant under the isometry generated by the spacetime vector $k$,
$\delta_{k}\varphi=0$. This implies, in particular, that $k$ is a Killing
vector and, for us, it will be the Killing vector associated to the Black
hole's Killing horizon. For the scalar fields it means that their Lie
derivatives with respect to that vector vanishes

\begin{equation}
  \delta_{k}\phi^{x}
  =
  -\pounds_{k}\phi^{x}
  =
  -\imath_{k}d\phi^{x}
  =
  0\,.
\end{equation}

As shown in \cite{Elgood:2020svt,Elgood:2020mdx,Elgood:2020nls},\footnote{The
  work terms for the electric charges associated to $p$-forms were found using
  the covariant phase space formalsm in Ref.~\cite{Compere:2007vx}. See also
  \cite{Prabhu:2015vua} for a different, equivalent, approach based on the
  mathematics of principal bundles. The importance of the gauge- and
  diffeomorphism invariance of the charges and potentials that occur in the
  laws of black-hole thermodynamics has been stressed in
  \cite{Hajian:2015xlp,Hajian:2022lgy} and in the 5th chapter of
  \cite{Grumiller:2022qhx}. } for the 1-form fields, it means that their Lie
derivatives with respect to $k$ plus a gauge transformation with parameter

\begin{equation}
  \chi_{k}
  =
  \imath_{k}A-P_{k}\,,
\end{equation}

\noindent
where the \textit{Maxwell momentum map} $P_{k}$ satisfies the \textit{Maxwell
  momentum map equation}\footnote{The local existence of a $P_{k}$ satisfying
  this equation follows from the assumption:
  \begin{equation}
    \delta_{k}F
    =
    -\pounds_{k}F
    =
    -d\imath_{k}F
    =
    0\,.
\end{equation}
}

\begin{equation}
  \label{eq:Maxwellmomentummapequation}
  \imath_{k}F+dP_{k}
  =
  0\,,
\end{equation}

\noindent
vanish identically:

\begin{equation}
  \begin{aligned}
    \delta_{k}A^{M} & =
    -\pounds_{k}A^{M} +d\chi_{k}^{M} =
    -\left(\imath_{k}d+d\imath_{k}\right)A^{M} +d\left(
      \imath_{k}A^{M}-P_{k}{}^{M}\right)
    \\
    & \\
    & =
    -\left( \imath_{k}F^{M}+dP_{k}{}^{M}\right) = 0\,,
  \end{aligned}
\end{equation}

\noindent
by virtue of the Maxwell momentum map equation
(\ref{eq:Maxwellmomentummapequation}).

If all the fields are invariant under $\delta_{k}$, so must the NGZ currents
be. Furthermore, since the NGZ currents are not gauge invariant, we must use
this definition for $\delta_{k}$:

\begin{equation}
  \begin{aligned}
    \delta_{k}\star j_{A}
    & =
    -\pounds_{k}\star j_{A} +\delta_{\chi_{k}}\star j_{A}
    \\
    & \\
    & =
    -\left(\imath_{k}d+d\imath_{k}\right)\star j_{A}
    -\tfrac{1}{2}\Omega_{MP}T_{A}{}^{P}{}_{N}\delta_{\chi_{k}}A^{M}\wedge F^{N}
    \\
    & \\
    & \doteq
    -d\imath_{k}\star j_{A}
    -\tfrac{1}{2}\Omega_{MP}T_{A}{}^{P}{}_{N}d\chi_{k}{}^{M}\wedge F^{N}
    \\
    & \\
    & \doteq
    d\left\{-\imath_{k}\star j_{A}
      -\tfrac{1}{2}\Omega_{MP}T_{A}{}^{P}{}_{N}\chi_{k}{}^{M} F^{N}\right\}
    \\
    & \\
    & =
    0\,,
  \end{aligned}
\end{equation}

\noindent
by assumption.

The expression in brackets is a 2-form that satisfies a Gauss law. Massaging
it a bit, we find the following manifestly gauge-invariant expression for it:

\begin{equation}
  \label{eq:charge2form}
  \mathbf{Q}_{A}[k]
  =
  \imath_{k}\star \hat{k}_{A}
    +\Omega_{MP}T_{A}{}^{P}{}_{N}P_{k}{}^{M}F^{N}\,.
\end{equation}

Now, integrating over 2-dimensional, spacelike, closed surfaces (and restoring
the normalization) we get the charges associated to the NGZ currents:

\begin{equation}
  Q_{A,k}
  =
  \frac{1}{16\pi G_{N}^{(4)}}
  \int_{\Sigma^{2}} \left\{\imath_{k}\star \hat{k}_{A}
    +\Omega_{MP}T_{A}{}^{P}{}_{N}P_{k}{}^{M}F^{N}\right\}\,.
\end{equation}

This is our proposal for scalar charges. Observe that under a duality
transformation generated by $k_{A},T_{A}$ with Lie brackets and commutation
relations

\begin{equation}
    [k_{A},k_{B}] = -f_{AB}{}^{C}k_{C}\,,
    \hspace{1cm}
    [T_{A},T_{B}] = +f_{AB}{}^{C}T_{C}\,,
\end{equation}

\noindent
these charges transform in the adjoint representation of the duality group:

\begin{equation}
\delta_{A}  Q_{B,k}
=
-f_{AB}{}^{C}Q_{C,k}\,.
\end{equation}

In what follows we are going to show in several examples corresponding to
static dilaton and axidilaton black holes that their values are non-vanishing
and reproduce the values of the conventionally-defined scalar charges
Eq.~(\ref{eq:conventionaldefinition}) but, before we set to do that, let us
observe that this definition depends on the value of the momentum map over the
integration surface. The Maxwell momentum map is defined only up to an
additive constant. This constant can be chosen so that
$\left.P_{k}{}^{M}\right|_{\infty}=0$. That is the choice that allows us to
recover the values of the conventionally-defined scalar charges
Eq.~(\ref{eq:conventionaldefinition}). However, other choices are
possible. The form of the first law that we are going to find includes an
additional term that takes into account that possibility so that the first law
is invariant under a change of asymptotic value of the Maxwell momentum maps.

It is also worth stressing that in the case in which we are considering (a
symmetric $\sigma$-model) there are always more symmetries than scalar
fields. Therefore, there are more 2-forms $\mathbf{Q}_{A}[k]$ satisfying a
Gauss law than scalars. Obviously, not all of them will be independent. In any
case, the conservation laws of those currents can be used to reconstruct the
equations of motion of the scalars using the identity

\begin{equation}
\delta_{x}{}^{y}= g^{AB}k_{A\, x}k_{B}{}^{y}\,,  
\end{equation}

\noindent
in which $g^{AB}$ is the Killing metric of the duality group $G$.

It also follows that there are more scalar charges than scalars, but we are
going to see that the conventionally-defined scalar charges $\Sigma^{x}$ can
be expressed in terms of the charges $Q_{A,k}$ that we have just defined.

It is worth mentioning that there is a slightly different procedure that
allows us to obtain the same expression Eq.~(\ref{eq:charge2form}) and that
was used in the case of dilaton black holes in Ref.~\cite{Pacilio:2018gom}. In
that case there is only one scalar and one target space Killing vector $k=1$
that generates the constant shifts of the scalar which are compensated by
rescalings of the vector field (see Section~\ref{sec-staticdilatonBHs}). In our
case, we have to project the scalar equations with the different Killing
vectors $k_{A}{}^{x}$ first, as in the first line of
Eq.~(\ref{eq:kAxEx}). Then, we take the inner product of the resulting
equation with $\imath_{k}$

\begin{equation}
    \imath_{k}k_{A}{}^{x}\mathbf{E}_{x}
    =
    -\imath_{k}d \star \hat{k}_{A}
    -\Omega_{MP}T_{A}{}^{P}{}_{N}\imath_{k}F^{M}\wedge F^{N}\,.
\end{equation}

If all the fields are invariant under the diffeomorphism generated by $k$

\begin{equation}
  \begin{aligned}
    -\imath_{k}d \star \hat{k}_{A}
    & =
    d\imath_{k} \star \hat{k}_{A}\,,
    \\
    & \\
    \imath_{k}F^{M}
    & =
    -dP_{k}{}^{M}\,,
  \end{aligned}
\end{equation}

\noindent
and, integrating by parts we arrive to $d\mathbf{Q}_{A}[k]=0$.

\section{First law and scalar charges}
\label{sec-firstlaw}

Taking into account the results obtained in
Refs.~\cite{Elgood:2020svt,Elgood:2020mdx,Elgood:2020nls,Ortin:2022uxa} for
the inclusion of matter fields, in Wald's formalism
\cite{Lee:1990nz,Wald:1993nt,Iyer:1994ys}, the first law of black hole
thermodynamics for a non-extremal black hole whose bifurcate horizon coincides
with the Killing horizon of the Killing vector field
$k=\partial_{t}+\Omega\partial_{\varphi}$, can be derived by integrating the
on-shell identity

\begin{equation}
  d\mathbf{W}[k]
  \doteq
  0\,,
\end{equation}

\noindent
where

\begin{equation}
  \mathbf{W}[k]
  \equiv
  \delta\mathbf{Q}[k] +\imath_{k}\mathbf{\Theta}(\varphi,\delta\varphi)-\varpi_{k}\,,
\end{equation}

\noindent
over a spacelike hypersurface with boundaries at spatial infinity
(S$^{2}_{\infty}$) and at the bifurcation sphere $\mathcal{BH}$ and applying
the Stokes theorem.

In the above identity $\mathbf{Q}[k]$ is the Noether-Wald charge for the
Killing vector $k$, $\mathbf{\Theta}(\varphi,\delta\varphi)$ is the
\textit{presymplectic $(d-1)$-form} defined in Ref.~\cite{Lee:1990nz} and
$\varpi_{k}$is defined by\footnote{This term arises when the effect of the
  induced gauge transformations are correctly taken into account as in
  Ref.~\cite{Ortin:2022uxa}. In Eq.~(\ref{eq:varpidef}) $\delta_{\Lambda_{k}}$
  stands for all the gauge transformations induced by the isometry generated
  by $k$.}

\begin{equation}
  \label{eq:varpidef}
  \delta_{\Lambda_{k}}\mathbf{\Theta}(\varphi,\delta\varphi)
  \equiv
  d\varpi_{k}\,.
\end{equation}

\noindent
Furthermore, it is assumed that the variations of the fields $\delta\varphi$
satisfy the linearized equations of motion in the black-hole's background.

The first law, thus, follows from the identity

\begin{equation}
  \int_{S^{2}{}_{\infty}}\mathbf{W}[k]
  =
  \int_{\mathcal{BH}}\mathbf{W}[k]\,.
\end{equation}

In previous works, following Ref.~\cite{Iyer:1994ys}, we assumed that, almost
by definition, the first integral simply gives the variation of the conserved
charges associated to the Killing vector $k$, that is,

\begin{equation}
\delta M-\Omega \delta J\,.
\end{equation}

A closer look reveals that, in presence of matter fields, it contains
additional terms that contribute to the first law \cite{Ortin:2022uxa}. In
particular, as we are going to see, it contains terms related to the scalar
charges that we have just defined. 

A standard calculation along the lines of
Refs.~\cite{Elgood:2020svt,Elgood:2020mdx,Elgood:2020nls,Ortin:2022uxa} gives

\begin{equation}
  \mathbf{Q}[k]
  =
  \star (e^{a}\wedge e^{b})P_{k\, ab} -P_{k}{}^{\Lambda}F_{\Lambda}\,,
\end{equation}

\noindent
where $P_{k\, ab}$ is the \textit{Lorentz momentum map} defined in
Ref.~\cite{Elgood:2020svt} and coincides with the \textit{Killing bivector}

\begin{equation}
P_{k\, ab} =\nabla_{a}k_{b}\,,  
\end{equation}

\noindent
and $P_{k}{}^{\Lambda}$ is the Maxwell momentum map defined in
Eq.~(\ref{eq:Maxwellmomentummapequation}). A quick calculation gives

\begin{equation}
  \label{eq:deltaQ}
  \delta\mathbf{Q}[k]
  =
  P_{k\, ab}\delta \star (e^{a}\wedge e^{b})
  +\star (e^{a}\wedge e^{b})\delta P_{k\, ab}
  -F_{\Lambda}\delta P_{k}{}^{\Lambda}
  -P_{k}{}^{\Lambda}\delta F_{\Lambda}\,.
\end{equation}

The presymplectic 3-form is given in Eq.~(\ref{eq:Theta}) and another short
calculation gives

\begin{equation}
  \label{eq:ikTheta}
  \begin{aligned}
    \imath_{k}\mathbf{\Theta}
    & =
    -\imath_{k}\star (e^{a}\wedge e^{b})\wedge
    \delta \omega_{ab} -\star (e^{a}\wedge e^{b})\wedge \delta\imath_{k}
    \omega_{ab} +g_{xy}\imath_{k}\star d\phi^{x}\delta\phi^{y}
    \\
    & \\
    & \hspace{.5cm}
    -\tfrac{1}{2}\imath_{k}F_{\Lambda}\wedge \delta A^{\Lambda}
    -\tfrac{1}{2}F_{\Lambda}\wedge \delta\imath_{k} A^{\Lambda}\,.
  \end{aligned}
\end{equation}

Since, on-shell, the dual 1-forms obey the same equations as the original
ones, we can define the \textit{dual (magnetic) momentum maps}
$P_{k\,\Lambda}$ through the equation

\begin{equation}
\imath_{k}F_{\Lambda}+dP_{k\,\Lambda}  =0\,,
\end{equation}

\noindent
and, substituting this definition in the above expression and integrating by
parts, we get

\begin{equation}
  \label{eq:ikTheta2}
  \begin{aligned}
    \imath_{k}\mathbf{\Theta}
    & =
    -\imath_{k}\star (e^{a}\wedge e^{b})\wedge
    \delta \omega_{ab} -\star (e^{a}\wedge e^{b})\wedge \delta\imath_{k}
    \omega_{ab} +g_{xy}\imath_{k}\star d\phi^{x}\delta\phi^{y}
    \\
    & \\
    & \hspace{.5cm}
    +P_{k\,\Lambda}\wedge \delta F^{\Lambda}
    -F_{\Lambda}\wedge \delta\imath_{k} A^{\Lambda}\,,
  \end{aligned}
\end{equation}

\noindent
up to an irrelevant total derivative.

Another simple calculation gives \cite{Ortin:2022uxa}

\begin{equation}
  \begin{aligned}
    \delta_{\Lambda_{k}} \mathbf{\Theta}
    & =
    (\delta_{\sigma_{k}}+\delta_{\chi_{k}}) \mathbf{\Theta}
    \\
    & \\
    & =
    -\delta_{\sigma_{k}}\left[\star (e^{a}\wedge e^{b})\wedge \delta
      \omega_{ab}\right]
    -F_{\Lambda}\wedge \delta_{\chi_{k}}\delta A^{\Lambda}
    \\
    & \\
    & =
    -\star (e^{a}\wedge e^{b})\wedge \mathcal{D}\delta \sigma_{k\, ab}
    -F_{\Lambda}\wedge d\delta \chi_{k}{}^{\Lambda}
    \\
    & \\
    & =
d\left\{    -\star (e^{a}\wedge e^{b})\wedge\delta \sigma_{k\, ab}
  -F_{\Lambda}\delta \chi_{k}{}^{\Lambda}
  \right\}\,,
  \end{aligned}
\end{equation}

\noindent
where the parameters of the induced Lorentz and Maxwell gauge transformations
are, respectively

\begin{subequations}
  \begin{align}
    \sigma_{k}{}^{ab}
    & =
      \imath_{k}\omega^{ab}-P_{k}{}^{ab}\,,
    \\
    & \nonumber \\
    \chi_{k}{}^{\Lambda}
    & =
      \imath_{k}A^{\Lambda}-P_{k}{}^{\Lambda}\,.
  \end{align}
\end{subequations}

Therefore,

\begin{equation}
  -\varpi_{k}
  =
  \star (e^{a}\wedge e^{b})\wedge\delta \sigma_{k\, ab}
  +F_{\Lambda}\delta \chi_{k}{}^{\Lambda}\,.
\end{equation}

Combining all these partial results, we arrive at

\begin{equation}
  \begin{aligned}
    \mathbf{W}[k]
    & =
      P_{k\, ab}\delta \star (e^{a}\wedge e^{b})
  -\imath_{k}\star (e^{a}\wedge e^{b})\wedge\delta \omega_{ab}
  \\
  & \\
  & \hspace{.5cm}
  -P_{k}{}^{\Lambda}\delta F_{\Lambda}
    +P_{k\,\Lambda}\delta F^{\Lambda}
  +g_{xy}\imath_{k}\star d\phi^{x}\delta\phi^{y}\,.
  \end{aligned}
\end{equation}

Let us consider the integral of $\mathbf{W}[k]$ at spatial infinity first,
restoring the global factor $1/(16\pi G_{N}^{(4)})$. The first two terms give
the gravitational contribution

\begin{equation}
\frac{1}{16\pi G_{N}^{(4)}}  \int_{S^{2}_{\infty}}\left\{
        P_{k\, ab}\delta \star (e^{a}\wedge e^{b})
  -\imath_{k}\star (e^{a}\wedge e^{b})\wedge\delta \omega_{ab}
\right\}
=
\delta M-\Omega\delta J\,,
\end{equation}

\noindent
while the third and fourth give\footnote{The electric and magnetic Maxwell
  momentum maps can be identified with the electrostatic and magnetostatic
  potentials $\Phi^{\Lambda}$ and $\Phi_{\Lambda}$, respectively.}

\begin{equation}
\frac{1}{16\pi G_{N}^{(4)}}  \int_{S^{2}_{\infty}}\left\{
  -P_{k}{}^{\Lambda}\delta F_{\Lambda}
    +P_{k\,\Lambda}\delta F^{\Lambda}
\right\}
=
  -\Phi^{\Lambda}_{\infty}\delta q_{\Lambda}
  +\Phi_{\Lambda\,\infty} \delta p^{\Lambda}
  =
  -\Omega_{MN}\Phi^{M}_{\infty}\delta q^{N}\,,
\end{equation}
  
\noindent
where $\Phi^{\Lambda}_{\infty}$ and $\Phi_{\Lambda\,\infty}$ are the values of
the electrostatic and magnetostatic potentials at spatial infinity.

Let us consider the last term. In the previous cases only conserved charges
are involved and it is natural to use the definition of scalar charges we have
proposed here to rewrite that term. Using the identity
$g_{xy}=g^{AB}k_{A\, x} k_{B\, y}$

\begin{equation}
  g_{xy}\imath_{k}\star d\phi^{x}\delta\phi^{y}
  =
  g^{AB} \imath_{k}\star \hat{k}_{A} k_{B\, y}\delta\phi^{y}
  =
  \left(\mathbf{Q}_{A}[k]-\Omega_{MP}T_{A}{}^{P}{}_{N}P_{k}{}^{M}F^{N}\right)\delta^{A}\,.
\end{equation}

\noindent
where we have defined

\begin{equation}
  \label{eq:deltaAdef}
\delta^{A} \equiv g^{AB}k_{B\, y}\delta\phi^{y}\,,  
\end{equation}

Restoring the global factor $1/(16\pi G_{N}^{(4)})$, we find

\begin{equation}
  \int_{S^{2}_{\infty}}
\left(\mathbf{Q}_{A}[k]-\frac{1}{16\pi G_{N}^{(4)}}\Omega_{MP}T_{A}{}^{P}{}_{N}P_{k}{}^{M}F^{N}\right)\delta^{A}
=
\left(\mathcal{Q}_{A\,k}
  -\Omega_{MP}T_{A}{}^{P}{}_{N}\Phi^{M}_{\infty}q^{N}\right)\delta^{A}_{\infty}\,.
\end{equation}

Then,

\begin{equation}
  \int_{S^{2}_{\infty}}\mathbf{W}[k]
  =
  \delta M-\Omega\delta J
  -\Omega_{MN}\Phi^{M}_{\infty}\delta q^{N}
  +\left(\mathcal{Q}_{A}
  -\Omega_{MP}T_{A}{}^{P}{}_{N}\Phi^{M}_{\infty}q^{N}\right)\delta^{A}_{\infty}\,.
\end{equation}

The bifurcation surface is defined by the property $k=0$ and, on it,

\begin{equation}
  P_{k\, ab}
  \stackrel{\mathcal{BH}}{=}
  \kappa n_{ab}\,,
\end{equation}

\noindent
where $n^{ab}$ is the binormal to the horizon wit the normalization
$n^{ab}n_{ab}=-2$ and $\kappa$ is the surface gravity. Therefore,

\begin{equation}
  \begin{aligned}
    \int_{\mathcal{BH}}\mathbf{W}[k]
    & =
    \frac{1}{16\pi G_{N}^{(4)}}
    \int_{\mathcal{BH}} \left\{P_{k\, ab}\delta \star (e^{a}\wedge e^{b})
      -P_{k}{}^{\Lambda}\delta F_{\Lambda} +P_{k\,\Lambda}\delta
      F^{\Lambda}\right\}
    \\
    & \\
    & =
    \frac{\kappa \delta A_{\mathcal{H}}}{8\pi G_{N}^{(4)}}
      -\Phi^{\Lambda}_{\mathcal{H}}\delta q_{\Lambda}
      +\Phi_{\Lambda\, \mathcal{H}}\delta p^{\Lambda}\,,
  \end{aligned}
\end{equation}

\noindent
where $A_{\mathcal{H}}$ is the area of the horizon and
$\Phi^{\Lambda}_{\mathcal{H}}$ and $\Phi_{\Lambda\,\mathcal{H}}$ are the
values of the electrostatic and magnetostatic potentials over the horizon
(constant according to the generalized zeroth law).

We arrive at our main result:\footnote{The overall sign of the electric and
  magnetic terms is unconventional. It is due to the definition of
  $F_{\Lambda}$ with a negative-definite kinetic matrix
  $I_{\Lambda\Sigma}$. It can be easily be changed, but the relative sign
  between the electric and magnetic terms can only be changed at the expense
  of losing explicit symplectic invariance.}

\begin{equation}
  \label{eq:firstlaw1}
  \delta M
=
\frac{\kappa \delta A_{\mathcal{H}}}{8\pi G_{N}^{(4)}}
+\Omega\delta J
  -\Omega_{MN}\left(\Phi^{M}_{\mathcal{H}}-\Phi^{M}_{\infty}\right)\delta q^{N}
  -\left(\mathcal{Q}_{A\, k}
  -\Omega_{MP}T_{A}{}^{P}{}_{N}\Phi^{M}_{\infty}q^{N}\right)\delta^{A}_{\infty}\,.
\end{equation}

In this expression the object $\delta^{A}_{\infty}$ is unusual, but it
just reflects the different forms in which the dualities of the theory can
modify the values of the moduli at infinity, which are also naturally
associated to the charges that we have defined.

The last term involving $\Phi^{M}_{\infty}$ is also unusual, but it has to be
there if we are going to allow for potentials which do not vanish at
infinity. In the examples that we are going to study explicitly,
$\Phi^{M}_{\infty}=0$ and the scalar charges take the expected
value. Furthermore, in that case, the scalar term can be brought to the form
found in Ref.~\cite{Gibbons:1996af} (up to the normalization of the charges):

\begin{equation}
  -\mathcal{Q}_{A\,k}\delta^{A}_{\infty}
  =
  -\mathcal{Q}_{A\, k}g^{AB}k_{B}{}^{x}_{\infty}g_{xy\, \infty}\delta \phi^{y}_{\infty}
  =
  -\tfrac{1}{4}\Sigma^{x}g_{xy\, \infty}\delta \phi^{y}_{\infty}\,,
\end{equation}

\noindent
where the scalar charges defined through the asymptotic expansions,
$\Sigma^{x}$ are related to the ones associated to the duality symmetries
$Q_{A}$ by

\begin{equation}
  \label{eq:SigmasversusQs}
  \Sigma^{x}
  =
  4\mathcal{Q}_{A}g^{AB}k_{B}{}^{x}_{\infty}\,.
\end{equation}

Finally, observe that, on the bifurcation surface

\begin{equation}
  \mathbf{Q}_{A}[k]
\stackrel{\mathcal{BH}}{=}
    \Omega_{MP}T_{A}{}^{P}{}_{N}P_{k}{}^{M}_{\mathcal{H}}F^{N}\,,
\end{equation}

\noindent
and, therefore

\begin{equation}
  \mathcal{Q}_{A\, k}
  =
  -\Omega_{MP}T_{A}{}^{P}{}_{N}\Phi^{M}_{\mathcal{H}}q^{N}\,.  
\end{equation}

This formula, which is our second main result, gives a universal relation
between the scalar charges of a black hole and the electric and magnetic
charges and potentials evaluated on the horizon generalizing the result found
in Ref.~\cite{Pacilio:2018gom} in a gauge-invariant way. Observe that The
existence of a bifurcate Killing horizon is crucial: in other spacetime
backgrounds the scalar charges may take arbitrary values in agreement with the
no-hair ``theorem'' and the interpretation of the non-trivial scalar fields of
these black-hole solutions as secondary scalar hair
\cite{Coleman:1991ku}.\footnote{Static, spherically-symmetric solutions of
  pure gravity and dilaton gravity with primary scalar hair
  (\textit{i.e.}~scalar fields with charges which are independent parameters
  of the solutions) can be found in Refs.~\cite{Janis:1968zz,Agnese:1994zx}
  (see also the higher-dimensional generalizations in Chapter~16 of
  Ref.~\cite{Ortin:2015hya}) and are singular.}

If we plug that formula back into the first law we arrive at

\begin{equation}
    \label{eq:firstlaw2}
  \delta M
=
\frac{\kappa \delta A_{\mathcal{H}}}{8\pi G_{N}^{(4)}}
+\Omega\delta J
  -\Omega_{MN}\left(\Phi^{M}_{\mathcal{H}}-\Phi^{M}_{\infty}\right)\delta q^{N}
  -\Omega_{MP}T_{A}{}^{P}{}_{N}\left(\Phi^{M}_{\mathcal{H}}-\Phi^{M}_{\infty}\right)q^{N}
  \delta^{A}_{\infty}\,,
\end{equation}

\noindent
which is manifestly independent of the choice of asymptotic value of the
potentials.

Notice that the right-hand side of this expression only contains the
variations of quantities which are independent physical parameters of the
black-hole solutions. The variations of the scalar charges cannot and do not
appear. The scalar charges actually pleay the roles of thermodynamical
potentials.

In the next two sections we are going to compare the scalar charges we have
defined with those obtained through the asymptotic expansion and the first law
that we have obtained with the first law obtained through the variation of the
entropy with respect to the physical parameters in two sets of solutions:
static, electrically-charged black holes and static axion-dilaton black holes.

\section{Static dilaton black hole solutions}
\label{sec-staticdilatonBHs}

Dilaton black holes are solutions of the family of models defined by the
action\footnote{Sometimes they are called Einstein-Maxwell-Dilaton (EMD)
  actions. We set $G_{N}^{(4)}=1$ throughout all this section.}

\begin{equation}
  \label{eq:actiondilatonBHs}
    S[e,A,\phi]
     =
      \frac{1}{16\pi}
      \int \left\{ -\star(e^{a}\wedge e^{b})
      \wedge R_{ab}
      +\tfrac{1}{2}d\phi\wedge \star d\phi
      +\tfrac{1}{2}e^{-a\phi}F\wedge \star F \right\}\,,
\end{equation}

\noindent
which depends on the real parameter $a$ and determines the strength of the
coupling of the dilaton and the Maxwell field. The static black-hole
solutions\footnote{Related solutions with primary scalar hair were found in
  Ref.~\cite{Agnese:1994zx}.}  of this model were found in
Refs.~\cite{Gibbons:1982ih,Gibbons:1984kp,Holzhey:1991bx} and can be written
in the form

\begin{equation}
  \begin{aligned}
    ds^{2}
    & =
    H^{-\frac{2}{1+a^{2}}}W dt^{2}
    -H^{\frac{2}{1+a^{2}}}\!\left[W^{-1}dr^{2} +r^{2}d\Omega_{(2)}^{2}
    \right]\,,
    \\
    & \\
    A_{t}
    & =
    \alpha e^{a\phi_{\infty}/2}(H^{-1}-1)\,,
    \\
    & \\
    e^{-\phi}
    & =
    e^{-\phi_{\infty}} H^{\frac{2a}{1+a^{2}}}\,,
  \end{aligned}
\end{equation}

\noindent
where the functions $H$ and $W$ take the form

\begin{equation}
  H = 1+\frac{h}{r}\,,
  \hspace{1cm}
  W = 1+\frac{\omega}{r}\,,
\end{equation}

\noindent
and the integration constants $h,\omega,\alpha$ satisfy the following
relation

\begin{equation}
\omega= h\!\left[1 -(1+a^{2})(\alpha/2)^{2} \right]\,.  
\end{equation}

In terms of the physical parameters $M,q,\phi_{\infty}$ (ADM mass, electric
charge and modulus) and the coupling constant $a$, the integration constants
$h,\omega$ and $\alpha$ are given by

\begin{equation}
  \begin{aligned}
    h
    & =
    -\frac{a^{2}+1}{a^{2}-1}\left\{M
      -\sqrt{M^{2}+4(a^{2}-1)e^{a\phi_{\infty}}q^{2}}\right\}\,,
    \\
    & \\
    \omega
    & =
    -\frac{2}{a^{2}-1}\left\{a^{2}M
      -\sqrt{M^{2}+4(a^{2}-1)e^{a\phi_{\infty}}q^{2}}\right\}\,,
    \\
    & \\
    \alpha
    & =
    -4qe^{a\phi_{\infty}/2}/h\,,
  \end{aligned}
\end{equation}

\noindent
for $a\neq 1$ and 

\begin{equation}
  \begin{aligned}
    h
    & =
\frac{4e^{\phi_{\infty}}q^{2}}{M}\,,
    \\
    & \\
    \omega
    & =
-2\frac{M^{2}-2e^{\phi_{\infty}}q^{2}}{M}\,,
    \\
    & \\
    \alpha
    & =
    -e^{-\phi_{\infty}/2}M/q\,.
  \end{aligned}
\end{equation}

The scalar charge $\Sigma$, computed using the conventional asymptotic
definition

\begin{equation}
  \label{eq:Sigmadefined}
\phi \sim \phi_{\infty}+\frac{\Sigma}{r}\,,  
\end{equation}

\noindent
takes the value

\begin{equation}
  \Sigma
  =
  -\frac{2ah}{a^{2}+1}\,.
\end{equation}

We have chosen the sign of the square roots in $h$ and $\omega$ so as to
always have $h>0$ and $\omega$ negative if certain non-extremality conditions
are met: for all values of $a$

\begin{equation}
    M^{2} > \frac{4}{a^{2}+1}e^{a\phi_{\infty}}q^{2}\,.
  \end{equation}

\noindent
In that case, there is an event horizon at

\begin{equation}
r=-\omega\equiv r_{0}\,,  
\end{equation}

\noindent
with Bekenstein-Hawking entropy

\begin{equation}
S =\pi r_{0}^{\frac{2a^{2}}{a^{2}+1}}(r_{0}+h)^{\frac{2}{a^{2}+1}}\,,  
\end{equation}

\noindent
and Hawking temperature

\begin{equation}
  T
  = \frac{r_{0}}{4S}\,.
\end{equation}

We can derive the first law for these families of black holes by varying the
entropy with respect to all the independent physical parameters, including the
modulus $\phi_{\infty}$:

\begin{equation}
    \delta S
     =
    \frac{1}{T}\left[\delta M
      +\frac{4}{(a^{2}+1)\alpha}e^{a\phi_{\infty}/2}\delta q
      +\tfrac{1}{4}\Sigma\delta\phi_{\infty}\right]\,,
\end{equation}

\noindent
for $a^{2}\neq 1$ and

\begin{equation}
    \delta S
     =
    \frac{1}{T}\left[\delta M
      +\frac{2}{\alpha}e^{\phi_{\infty}/2}\delta q
      +\tfrac{1}{4}\Sigma \delta\phi_{\infty}\right]\,,
\end{equation}

\noindent
for $a^{2}=1$.

In the above expressions $\Sigma$ is the scalar charge defined through the
asymptotic expansion Eq.~(\ref{eq:Sigmadefined}).

These theories are invariant under the global transformations generated by 

\begin{equation}
  \delta \phi = -1\,,
  \hspace{1cm}
  \delta A = -\frac{a}{2}A\,,
\end{equation}

\noindent
and Eq.~(\ref{eq:charge2form}) takes the form

\begin{equation}
  \mathbf{Q}[k]
  =
-\frac{1}{16\pi}\left\{  \imath_{k}\star d\phi
  +\frac{a}{2}P_{k}e^{-a\phi}\star F\right\}
=
-\frac{ah}{8\pi(a^{2}+1)}\omega_{(2)}\,,
\end{equation}

\noindent
where $\omega_{(2)}$ is the volume form of the round 2-sphere of unit radius.
It is evident that these 2-forms satisfy a Gauss law and they give the same
value when they are integrated over 2-spheres of any radius:

\begin{equation}
  \mathcal{Q}_{k}
  =
  -\frac{ah}{2(a^{2}+1)}
  =
  -\tfrac{1}{4}\Sigma\,,
\end{equation}

\noindent
as expected according to our general arguments. This is, essentially, the
result obtained by Pacilio in Ref.~\cite{Pacilio:2018gom}.

\section{Static axion-dilaton black hole solutions}
\label{sec-staticBHs}

The so-called axion-dilaton model is just a generalization to an arbitrary
number of vector fields $n_{V}$ of pure, ungauged, $\mathcal{N}=4,d=4$
supergravity \cite{Cremmer:1977tt}, although this model can also be embedded
in $\mathcal{N}=2,d=4$ supergravity for $n_{V}=2$.

We can introduce it as a model with two real scalars $\phi^{1}=a$ (the axion)
and $\phi^{2}=\phi$ (the dilaton) which are naturally combined into the
complex scalar (\textit{axidilaton})

\begin{equation}
\lambda = a+i e^{-2\phi}\,,   
\end{equation}

\noindent
and where the $\sigma$-model metric and the period matrix are given by

\begin{equation}
  (g_{xy})
  =
  \left(
    \begin{array}{lr}
     e^{4\phi} & 0 \\ 0 & 4 \\
    \end{array}
  \right)\,,
  \hspace{1cm}
  \mathcal{N}_{\Lambda\Sigma}
  =
  -\lambda \delta_{\Lambda\Sigma}\,. 
\end{equation}

The most general non-extremal, static, black-hole solution of the
axion-dilaton model was presented in Ref.~\cite{Kallosh:1993yg} and it is a
generalization of the solutions presented in
Refs.~\cite{Gibbons:1982ih,Gibbons:1987ps,Garfinkle:1990qj,Shapere:1991ta,Kallosh:1992ii,Ortin:1992ur}.\footnote{The
  most general stationary, non-extremal black-hole solution of this theory was
  presented in Ref.~\cite{LozanoTellechea:1999my}.} A very useful feature of
this solution is that it is written in terms of its physical parameters only:
the ADM mass $M$, the asymptotic value of the axidilaton
$\lambda_{\infty}=a_{\infty}+ie^{-2\phi_{\infty}}$, the complex
electromagnetic charges $\Gamma^{\Lambda}$ (a combination of the real electric
charges $q_{\Lambda}$, the real magnetic charges $p_{\Lambda}$ and the moduli
$\lambda_{\infty}$) and the complex axidilaton charge
$\Upsilon=\Sigma+i\Delta$. All these parameters are defined by the asymptotic
expansions ($G_{N}^{(4)}=1$)

\begin{subequations}
  \begin{align}
    g_{tt}
    & \sim  1-\frac{2M}{r}\,,
    \\
    & \nonumber \\
    \lambda
    & \sim
    \lambda_{\infty}-ie^{-2\phi_{\infty}} \frac{2\Upsilon}{r}\,,
    \\
    & \nonumber \\
    \tfrac{1}{2}\left[F^{\Lambda}{}_{tr}+i\star F^{\Lambda}{}_{tr}\right]
    & \sim
    \frac{e^{+\phi_{\infty}}\Gamma^{\Lambda}}{r^{2}}
    =\frac{e^{+2\phi_{\infty}}(q_{\Lambda}-\lambda_{\infty}^{*}p^{\Lambda})}{r^{2}}\,.
  \end{align}
\end{subequations}

The asymptotic behavior of $\lambda$ implies for those of $a$ and $\phi$

\begin{subequations}
  \begin{align}
    a
    & \sim
      a_{\infty}+\frac{2e^{-2\phi_{\infty}}\Im\mathfrak{m}\Upsilon}{r}\,,
    \\
    & \nonumber \\
    \phi
    & \sim
      \phi_{\infty}+\frac{\Re\mathfrak{e}\Upsilon}{r}\,,
  \end{align}
\end{subequations}

\noindent
so

\begin{equation}
  \label{eq:asymptoticscalarchargesaxidilaton}
  \Sigma^{1}
  =
  2e^{-2\phi_{\infty}}\Im\mathrm{m}(\Upsilon)\,,
  \hspace{1cm}
  \Sigma^{2}
  =
  \Re\mathrm{e}(\Upsilon)\,.
\end{equation}

\noindent
The axidilaton charge is a function of the rest of the physical parameters:

\begin{equation}
  \label{eq:upsilongamma}
  \Upsilon
  =
  -\frac{2}{M}\Gamma^{\Lambda\, *}\Gamma^{\Lambda\, *}\,.
\end{equation}

The ADM mass can be defined more rigorously as a conserved quantity through
the ADM \cite{Arnowitt:1959ah}, the Abbott-Deser \cite{Abbott:1981ff} or many
other formalisms. The electric and magnetic charges can also be defined as
conserved charges by standard methods
Refs.~\cite{Barnich:2001jy,Barnich:2003xg} as

\begin{subequations}
  \begin{align}
    p^{\Lambda}
    & \equiv 
      \frac{1}{16\pi G_{N}^{(4)}}\int F^{\Lambda}\,,
    \\
    & \nonumber \\
    q_{\Lambda}
    & \equiv 
      \frac{1}{16\pi G_{N}^{(4)}}\int F_{\Lambda}\,.
  \end{align}
\end{subequations}

In contrast, as we have stressed, the scalar charges are conventionally
defined through the above asymptotic expansion which is not based on any
conservation (\textit{Gauss}) law. Our goal in this section will be to show
that the definition of scalar charges that we have proposed in
Section~\ref{sec-scalarcharge} gives exactly the same result for the static
solutions of the axion-dilaton model.

The most economical way of presenting this kind of solutions is through the
time components of the original and dual 1-form fields $A^{\Lambda}{}_{t}$ and
$A_{\Lambda\, t}$, respectively. They contain enough information to recover
the rest of the components of each of them.\footnote{They are computed
  explicitly in Ref.~\cite{Mitsios:2021zrn}.} The solution is, then,
\cite{Kallosh:1993yg}

\begin{equation}
  \label{eq:axidilatonsolutions}
  \begin{aligned}
    ds^{2}
    & =
    e^{2U}dt^{2}-e^{-2U}dr^{2}-R^{2}d\Omega^{2}_{(2)}\,,
    \\
    & \nonumber \\
    \lambda
    & = 
    \frac{\lambda_{\infty}r+\lambda^{*}_{\infty}\Upsilon}{r+\Upsilon}\,,
    \\
    & \\
    A^{\Lambda}{}_{t}
    & = 
    2e^{\phi_{\infty}}R^{-2}[\Gamma^{\Lambda}(r+\Upsilon)+\mathrm{c.c.}]\,,
     \\
    & \\
    A_{\Lambda\, t}
    & = 
    -2e^{\phi_{\infty}}R^{-2}
    [\Gamma^{\Lambda}(\lambda_{\infty}r+\lambda^{*}_{\infty} \Upsilon)
    +\mathrm{c.c.}]\,. 
  \end{aligned}
\end{equation}

\noindent
The functions $e^{2U}$ and $R$ are given by

\begin{equation}
  \begin{aligned}
    e^{2U}
    & =
    R^{-2}(r-r_{+})(r-r_{-})\,,
    \\
    & \\
    R^{2} & =
    r^{2}-|\Upsilon|^{2}\,,
  \end{aligned}
\end{equation}

\noindent
and the parameters $r_{\pm}$ that appear in $e^{2U}$ (actually, the positions
of the outer and inner horizons when they take real values, \textit{i.e.}~when
$r_{0}^{2}> 0$) are given by

\begin{equation}
  r_{\pm}    = M\pm r_{0}\,,
  \,\,\,\,\,
  \text{with}
  \,\,\,\,\,
  r_{0}^{2}=M^{2}+|\Upsilon|^{2}-4\Gamma^{\Lambda} \Gamma^{\Lambda\,*}\,.
\end{equation}

Since we are just interested in the thermodynamics of these black holes, we
only need their Hawking temperature and Bekenstein-Hawking entropy, which are
given by 

\begin{subequations}
  \begin{align}
    T
    & =
      \frac{r_{0}}{2S}\,,
    \\
    & \nonumber \\
    S
    & =
      2\pi\left\{M^{2}+Mr_{0}-2\Gamma^{\Lambda\, *}\Gamma^{\Lambda\, *}\right\}\,.
  \end{align}
\end{subequations}

Varying $S$ with respect to the physical charges $M,q_{\Lambda},p^{\Lambda}$
and the moduli $\lambda_{\infty}$ we get the first law:

\begin{equation}
  \delta M = T\delta S +\Phi^{\Lambda}\delta q_{\Lambda}
  -\Phi_{\Lambda}\delta p^{\Lambda}
  -\tfrac{1}{2}\Im\mathrm{m}(\Upsilon)e^{2\phi_{\infty}}\delta a_{\infty}
  -\Re\mathrm{e}(\Upsilon) \delta\phi_{\infty}\,.
\end{equation}

The last two terms can be rewritten in two different fashions:

\begin{equation}
  \begin{aligned}
    -\tfrac{1}{2}\Im\mathrm{m}(\Upsilon)e^{2\phi_{\infty}}\delta a_{\infty}
    -\Re\mathrm{e}(\Upsilon) \delta\phi_{\infty}
    & =
    -\tfrac{1}{2}\Im\mathrm{m}
    \left(\Upsilon^{*}\frac{\delta \lambda_{\infty}}{e^{-2\phi_{\infty}}}
    \right)
    \\
    & \\
    & = -\tfrac{1}{4} g_{xy}(\phi_{\infty})\Sigma^{x}\delta\phi^{y}_{\infty}\,,
  \end{aligned}
\end{equation}

\noindent
where $\Sigma^{1},\Sigma^{2}$ are the asymptotic scalar charges defined in
Eqs.~(\ref{eq:asymptoticscalarchargesaxidilaton}).  Both expressions are
manifestly duality-invariant.\footnote{$e^{2\phi}\delta\lambda$ and $\Upsilon$
  are multiplied by the same phase under SL$(2,\mathbb{R})$ transformations.}

We are now going to see how the scalar charges $\Sigma^{x}$ are related to
those defined in Section~\ref{sec-scalarcharge} and how the scalar term in the
first law agrees with the one in Eq.~(\ref{eq:firstlaw1}).

The Killing vectors of the target-space metric are

\begin{equation}
  k_{1}=a\partial_{a} -\tfrac{1}{2}\partial_{\phi}\,,
  \hspace{.5cm}
  k_{2}= \tfrac{1}{2}(1-a^{2}+e^{-4\phi})\partial_{a}
  +\tfrac{1}{2}a\partial_{\phi}\,,
  \hspace{.5cm}
  k_{3}= \tfrac{1}{2}(1+a^{2}-e^{-4\phi})\partial_{a}
  -\tfrac{1}{2}a\partial_{\phi}\,,  
\end{equation}

\noindent
and their Lie brackets satisfy the sl$(2,\mathbb{R})\sim$so$(2,1)$ algebra

\begin{equation}
\left[k_{A},k_{B}\right] = \varepsilon_{ABD}\eta^{DC}k_{C}\,,  
\end{equation}

\noindent
where $(\eta_{AB})=(\eta^{AB})=$diag$(++-)$ is the SO$(2,1)$ invariant metric.

The SL$(2,\mathbb{R})$ matrices which act on the 1-form fields are tensor
products 

\begin{equation}
  S
  =
  \left(
    \begin{array}{cc}
    A  & B \\ C & D \\
    \end{array}
  \right)\otimes \mathbb{1}_{n_{V}\times n_{V}}\,,
  \,\,\,\,\,\,
  \text{with}
  \,\,\,\,\,\,
  AD-BC=1\,.
\end{equation}

The generators (always $\otimes \mathbb{1}_{n_{V}\times n_{V}}$) are

\begin{equation}
  T_{1} = -\tfrac{1}{2}\sigma^{3}\,,
  \hspace{1cm}
  T_{2} = -\tfrac{1}{2}\sigma^{1}\,,
  \hspace{1cm}
  T_{3} = \tfrac{i}{2}\sigma^{2}\,,
\end{equation}

\noindent
and their commutation relations are

\begin{equation}
\left[T_{A},T_{B}\right] = -\varepsilon_{ABD}\eta^{DC}k_{C}\,.  
\end{equation}

It is somewhat simpler to work with the bases $k_{1},k_{\pm}=k_{2}\pm k_{3}$
and $T_{1},T_{\pm}=T_{2}\pm T_{3}$. We compute separately the
$\imath_{k}\star \hat{k}_{A}$ and $\Omega_{MP}T_{A}{}^{P}{}_{N}P_{k}^{M}F^{N}$
contributions, which in this case correspond to

\begin{equation}
  \begin{aligned}
    \Omega_{MP}T_{1}{}^{P}{}_{N}P_{k}^{M}F^{N}
    & =
    \tfrac{1}{2}\left(P_{k}^{\Lambda}F_{\Lambda}+P_{k\,
        \Lambda}F^{\Lambda}\right)\,,
    \\
    & \\
    \Omega_{MP}T_{+}{}^{P}{}_{N}P_{k}^{M}F^{N}
    & =
    -P_{k}^{\Lambda}F^{\Lambda}\,,
    \\
    & \\
    \Omega_{MP}T_{-}{}^{P}{}_{N}P_{k}^{M}F^{N}
    & =
    P_{k\, \Lambda}F_{\Lambda}\,.
  \end{aligned}
\end{equation}

In this case we can use as potentials the time components of the original and
dual vector fields given in Eqs.~(\ref{eq:axidilatonsolutions})

\begin{equation}
P_{k}^{M} = A^{M}{}_{t}\,,  
\end{equation}

\noindent
which vanish identically at infinity.

We are only interested in the pullback of these 2-forms over
2-spheres.\footnote{There are additional $tr$ components that we ignore since they
  do not contribute to the integrals.} The results, after a long calculation
are ($G_{N}^{(4)}=1$)

\begin{equation}
  \begin{aligned}
    \mathbf{Q}_{1\,k}
    & =
    -\frac{1}{8\pi} e^{2\phi_{\infty}}
    \Im\mathfrak{m}(\lambda_{\infty}^{*}\Upsilon)\omega_{(2)}\,,
    \\
    & \\
    \mathbf{Q}_{+\,k}
    & =
    -\frac{1}{8\pi} e^{2\phi_{\infty}}
    \Im\mathfrak{m}(\Upsilon)\omega_{(2)}\,,
    \\
    & \\
    \mathbf{Q}_{-\,k}
    & =
    \frac{1}{8\pi} e^{2\phi_{\infty}}
    \Im\mathfrak{m}(\lambda_{\infty}^{*\,2}\Upsilon)\omega_{(2)}\,,
    \\
    & \\
  \end{aligned}
\end{equation}

Again, it is evident that these 2-forms satisfy a Gauss law and they give the same
value when they are integrated over 2-spheres of any radius, namely

\begin{equation}
  \begin{aligned}
    \mathcal{Q}_{1\,k}
    & =
    -\tfrac{1}{2} e^{2\phi_{\infty}}
    \Im\mathfrak{m}(\lambda_{\infty}^{*}\Upsilon)\,,
    \\
    & \\
    \mathcal{Q}_{+\,k}
    & =
    -\tfrac{1}{2} e^{2\phi_{\infty}}
    \Im\mathfrak{m}(\Upsilon)\,,
    \\
    & \\
    \mathcal{Q}_{-\,k}
    & =
    \tfrac{1}{2} e^{2\phi_{\infty}}
    \Im\mathfrak{m}(\lambda_{\infty}^{*\,2}\Upsilon)\,.
  \end{aligned}
\end{equation}

It is now trivial to see that the asymptotic charges
Eqs.~(\ref{eq:asymptoticscalarchargesaxidilaton}) are recovered using
Eq.~(\ref{eq:SigmasversusQs}) with the Killing vectors given above and

\begin{equation}
  \left(g^{AB}\right)
  =
  \left(
    \begin{array}{ccc}
      1 & 0 & 0   \\
      0 & 0 & 1/2 \\
    1/2 & 0 & 0   \\
    \end{array}
  \right)\,,
\end{equation}
 
\noindent
for the $1,+,-$ basis.

The first law Eq.~(\ref{eq:firstlaw1}) is recovered with

\begin{equation}
  \delta_{1\,\infty}
  =
  \tfrac{1}{2}e^{4\phi_{\infty}}\delta|\lambda|^{2}_{\infty}\,,
\hspace{1cm}
  \delta_{+\,\infty}
  =
  e^{4\phi_{\infty}}\delta a_{\infty}\,,
  \hspace{1cm}
  \delta_{-\,\infty}
  =
  -\tfrac{1}{2}e^{4\phi_{\infty}}
  (\lambda^{*\,2}_{\infty}\delta\lambda_{\infty}+\mathrm{c.c.})\,.
\end{equation}

\section{Discussion}
\label{sec-discussion}

Some final comments on our results are in order.

First of all, it is unclear how to give coordinate-independent definitions of
scalar charges satisfying a Gauss law in absence of global symmetries. This
limitation led us to focus on theories with enough global symmetries to account
for all the possible scalar charges. On the other hand, there are not many
examples of black-hole solutions in theories with no or very few global
isometries. Most of the general recipes elaborated to construct black holes in
$\mathcal{N}=2,d=4$ theories, for instance, \cite{Meessen:2006tu} are only
valid for extremal black holes, which lie outside the scope of our methods.
Working with non-extremal black holes is much more difficult
\cite{Galli:2011fq} although some general methods have been developed
\cite{Meessen:2011aa} and they should be revisited to study this problem.

In general, a Gauss law is not equivalent to a full conservation law. In our
case, the restriction to backgrounds with timelike Killing vectors makes it
trivially equivalent to a conservation law in those particular backgrounds,
but not in general. We expect, however, that the existence of a rigorous
definition can be used to study the evolution of scalar charge or at least its
behavior under perturbations.

It is worth stressing the relation between the value of the scalar charge and
the existence of a regular bifurcate Killing horizon. In absence of such a
horizon there does not seem to be a restriction on that value. It is because
of this relation that it can be understood as secondary black-hole hair.

The general procedure that has allowed us to define a $(d-2)$-form satisfying
a Gauss law starting from the $(d-1)$-form (Noether current) associated to a
global symmetry can probably used in more general settings (fermionic matter,
for instance).

As we mentioned before, it should be stressed that these results can
be generalized to higher-rank fields and higher dimensions. The NGZ currents
have been determined in Ref.~\cite{Bandos:2016smv} and one simply has to
follow the same steps. It also seems that it should also be possible to find
$(d-2)$-forms satisfying Gauss laws starting from any standard Noether current
$(d-1)$-forms associated to a global symmetry.
            
Concerning the first law, in order to recover the GKK scalar term it has been
essential to realize that the integral of $\mathbf{W}[k]$ at spatial infinity
gives more than just the variations of the gravitational charges at
infinity. Often, these contributions have been ignored or set to zero via
convenient boundary conditions at spatial infinity. Often, the integral on the
bifurcation surface has been also identified with the $T\delta S$ term of the
firm law ignoring other contributions (work terms). We think it is now clear
that there are different contributions to the first law coming from that
integral as well and that the only one which is associate to the entropy is
the one that takes the form of a conserved Lorentz charge, as we have pointed
out in Refs.~\cite{Elgood:2020mdx,Elgood:2020nls,Ortin:2022uxa}. Actually, the
title of Ref.~\cite{Wald:1993nt} should be replaced by ``Black hole entropy is
the (Lorentz) Noether charge.''

\section*{Acknowledgments}

This work has been supported in part by the MCI, AEI, FEDER (UE) grants
PID2021-125700NB-C21 (``Gravity, Supergravity and Superstrings'' (GRASS)) and
IFT Centro de Excelencia Severo Ochoa CEX2020-001007-S. The work of RB has
also been supported by the National Agency for Research and Development [ANID]
Chile, Doctorado Nacional, under grant 2021-21211461 and by PUCV, Beca
Pasant\'{\i}a de Investigaci\'on. The work of CG-F was supported by the MU grant
FPU21/02222. The work of MZ was supported by the fellowship
LCF/BQ/DI20/11780035 from ``La Caixa'' Foundation (ID 100010434). TO wishes to
thank M.M.~Fern\'andez for her permanent support.


\end{document}